\documentclass[%
 reprint,
superscriptaddress,
 amsmath,amssymb,
 aps,prl
]{revtex4-2}
\usepackage{graphicx}
\usepackage{dcolumn}
\usepackage{bm}

\graphicspath{{fig/}}
\begin{document}

\preprint{APS/123-QED}

\title{Nucleation instability preempts relativistic domain wall transport in high-exchange ferrimagnetic nanowires}
\author{Pietro Diona}
\email{pietro.diona@sns.it}
\affiliation{Nanoscience, Scuola Normale Superiore, Piazza dei Cavalieri 7, Pisa, 56126, Italy}
\affiliation{Quantum Materials Theory, Italian Institute of Technology, Via Morego 30, Genoa, 16163, Italy}%
\affiliation{Department of Materials, ETH Zürich, Hönggerbergring 64, 8093 Zürich, Switzerland}

\date{\today}
\begin{abstract}
Current-driven domain wall motion in ferrimagnets can approach the spin-wave velocity, giving rise to relativistic-like dynamics. While this regime has been experimentally observed in crystalline ferrimagnetic garnets and amorphous GdFeCo, the material conditions that determine whether it can be accessed remain unresolved. Here, we investigate spin-orbit-torque-driven domain wall motion in high-exchange GdCo nanowires using magneto-optical Kerr effect microscopy. We find that nucleation preempts relativistic transport. In GdCo, no velocity saturation or high-field collapse is observed. Instead, the large exchange interaction raises the maximum spin-wave group velocity to $\approx 7\text{--}9~\mathrm{km/s}$, far above the experimentally accessible domain wall velocities. Before this limit can be approached, increasing current density and in-plane magnetic field induce domain nucleation, disrupting steady-state propagation. We map the boundary separating domain wall transport from nucleation instability and show that the nucleation threshold decreases with pulse duration, consistent with thermally assisted barrier crossing. These results identify nucleation as the mechanism that prevents access to the relativistic regime in high-exchange ferrimagnets and establish a dynamical phase boundary between steady propagation and nucleation.
\end{abstract}

\keywords{domain wall, ferrimagnet, magnon speed, spin-orbit torque, nucleation}



\maketitle

\paragraph{Introduction ---}

\begin{figure}[t]
     \centering
         \includegraphics[width=1\linewidth]{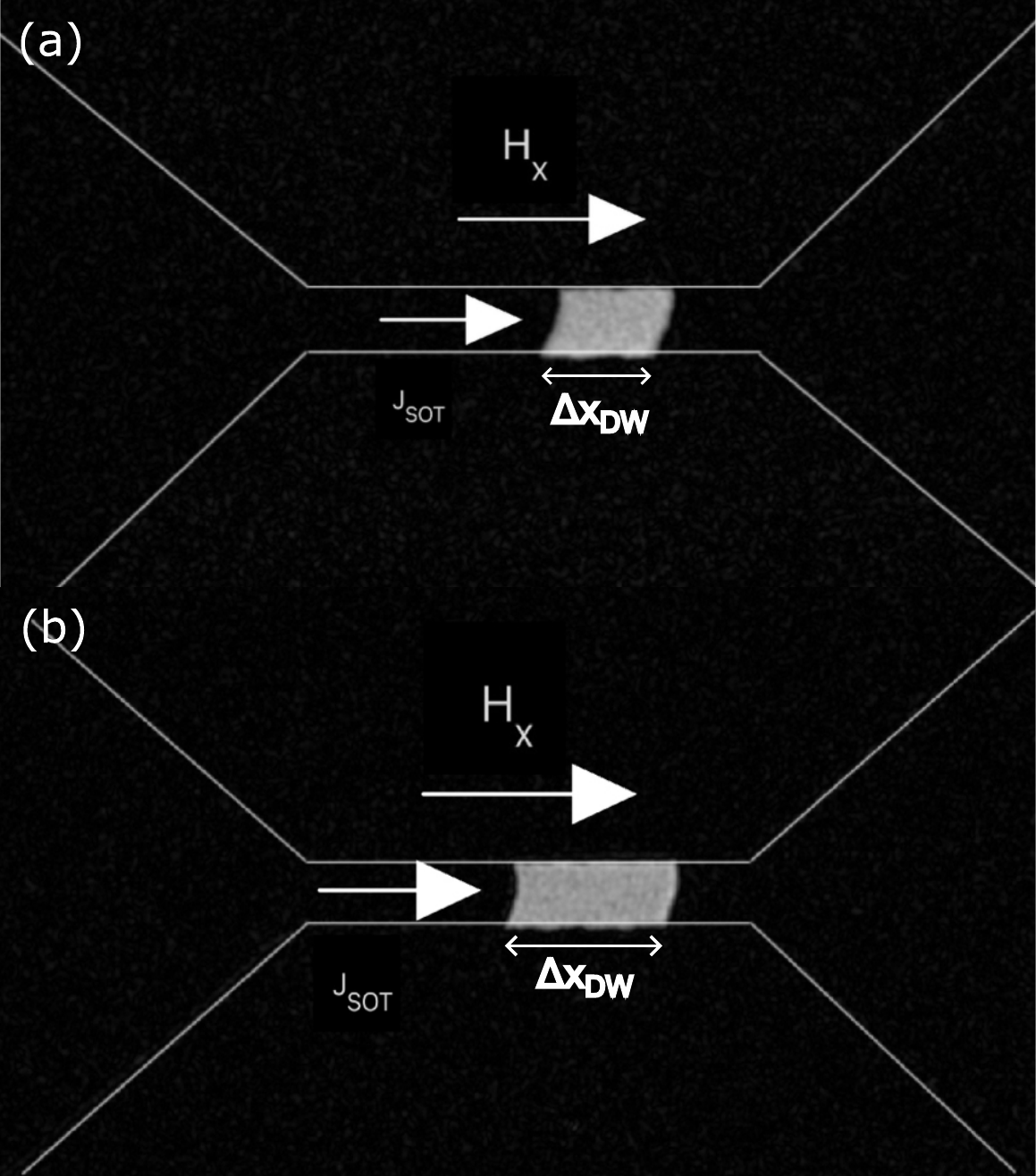}
    \caption{MOKE images of the displacement of an up-down domain wall after applying ten 3-ns-long pulses of $1.4\cdot 10^{12} \mathrm{A/m^2}$ under: (a) $H_\mathrm{x} = 11 \, \mathrm{mT}$, (b) $H_\mathrm{x} = 38 \, \mathrm{mT}$. The white contrast corresponds to the domain wall displacement, quantified by the displacement length $\Delta x_{\mathrm{DW}}$. Current density $j$ and in-plane magnetic field $B_x$ are applied along the nanowire axis.}
       \label{fig1}
\end{figure}

Magnetic domain walls are promising building blocks for high-speed spintronic devices and racetrack memories \cite{parkin2008magnetic}. Early control relied on spin transfer torque, where angular momentum is transferred between conduction electrons and local moments \cite{ralph2008spin, STT, brataas2012current}. Spin-orbit torques (SOTs) offer faster and more versatile control via the spin Hall or Rashba-Edelstein effects at adjacent layers or interfaces \cite{manchon2019current, ryu2020current, ramaswamy2018recent, SOT1}. While ferromagnets were initially explored \cite{beach2005dynamics, mougin2007domain, diona2022}, attention has progressively shifted toward antiferromagnets \cite{baldrati2019mechanism, gomonay2016high, baltz2018antiferromagnetic, shiino2016antiferromagnetic} and ferrimagnets \cite{caretta2024domain, kim2022ferrimagnetic, vsteady1, sala2022asynchronous, sala2023deterministic}, which combine reduced field sensitivity with ultrafast spin dynamics. Beyond high-speed memory and logic applications, current-driven magnetic textures have recently attracted interest as physical platforms for nonlinear spintronic computing, where pinning, stochasticity, and threshold activation can supply device-level nonlinear responses \cite{godinho2024antiferromagnetic, liu2024domain, marrows2024neuromorphic}.

In ferrimagnets, domain walls behave as relativistic sine-Gordon solitons and can approach the spin-wave group velocity, the maximum speed allowed for magnetic excitations, analogous to the speed of light in special relativity \cite{einstein1905elektrodynamik, Maranzana25, diona2026rocket}. This relativistic limit has been experimentally observed in crystalline BiYIG \cite{caretta2020relativistic} and in amorphous rare-earth--transition-metal (RE-TM) GdFeCo \cite{diona2025observation}. In GdFeCo, the relatively low exchange stiffness places $v_{\mathrm{g,max}}$ near $\sim 1\text{--}2~\mathrm{km/s}$, so SOT-driven domain walls can reach the relativistic regime, exhibit velocity saturation \cite{diona2025observation}. This raises a central question for ferrimagnetic spintronics: is relativistic transport a generic experimentally accessible regime, or does it compete with other phenomena?

Here, we address this question by investigating SOT-driven domain wall dynamics in GdCo nanowires. GdCo provides the high-exchange counterpar to GdFeCo: the estimated exchange stiffness $A \approx 10\text{--}12~\mathrm{pJ/m}$ raises $v_{\mathrm{g,max}}$ to $\approx 7\text{--}9~\mathrm{km/s}$, far above the velocities reached before nucleation instability sets in. We show that increasing current density and in-plane magnetic field induce spontaneous domain nucleation before the spin-wave velocity can be approached. This boundary  separates domain wall transport from a  nucleation regime at a threshold $j_{\mathrm{th}}\left(B_x\right)$ which decreases with pulse duration, consistent with thermally assisted barrier crossing. The result is not simply a failure to reach the relativistic regime: it identifies the competing mechanism that prevents access to relativistic transport in high-exchange ferrimagnets. In this sense, relativistic propagation and nucleation are distinct dynamical regimes of ferrimagnetic domain wall motion.

\paragraph{Domain wall motion in a ferrimagnetic nanowire ---}

We consider a ferrimagnet/heavy-metal heterostructure, where SOTs arise from the adjacent layer. Therefore, domain walls are driven by the combined action of the damping-like torque $\tilde{H}_\mathrm{DL} \sim \frac{\theta_{\mathrm{DL}} j}{M_{\mathrm{S}}}$ (proportional to the current density $j$ and damping-like SOT efficiency $\theta_{\mathrm{DL}}$ and inversely proportional to the saturation magnetization $M_{\mathrm{S}}$) and an in-plane magnetic field $H_\mathrm{x} = B_\mathrm{x}/\mu_0$ applied along the current direction \cite{manchon2019current}. The in-plane field, together with the Dzyaloshinskii–Moriya interaction $H_\mathrm{D}$, stabilizes a Néel configuration, enabling efficient current-driven domain wall motion \cite{khvalkovskiy2013matching}. In this regime, the steady-state dynamics follow a relativistic-like expression for the domain wall velocity $v_\mathrm{DW}$ \cite{caretta2020relativistic, diona2025observation}:
\begin{equation}
\label{rel2}
    v_\mathrm{DW} = \frac{v_\mathrm{g, max}}{\sqrt{1 + \left(\frac{v_\mathrm{g, max}}{v_\mathrm{cl}}\right)^{2}}},
\end{equation}
where $v_\mathrm{g, max} = \frac{2A}{ds_\mathrm{T}}$ is the maximum spin-wave group velocity. In Eq.~\ref{rel2}, $A$ is the antiferromagnetic exchange interaction, $d$ is the average interatomic distance , and $s_\mathrm{T}$ is the total spin density of the two sublattices composing the ferrimagnet. $v_\mathrm{cl}$ denotes the classical domain wall velocity in the low-velocity regime, i.e., well below $v_\mathrm{g,max}$ \cite{haltz2021domain, diona2025observation}:
\begin{eqnarray}
\label{vsteadycomplete}
    &v_\mathrm{cl} = \frac{\Delta\gamma_\mathrm{eff}\tilde{H}_\mathrm{pin}}{\alpha_\mathrm{eff}\beta}\left[1 - \sqrt{1 - \beta\left[1 - \left(\frac{2\eta\tilde{H}_\mathrm{DL}}{\tilde{H}_\mathrm{pin}}\right)^2\right]}\right],\nonumber\\
    & \beta = 1 + \left[ \frac{\frac{\pi}{2}\tilde{H}_\mathrm{DL}}{\alpha_\mathrm{eff}\left(H_\mathrm{D} + \frac{\pi}{2}H_\mathrm{x}\right)}\right]^{2}.
\end{eqnarray}
Here, $\Delta \sim \sqrt{\frac{A}{K_{\mathrm{eff}}}}$ is the domain wall width, $K_{\mathrm{eff}}$ is the effective anisotropy, $\alpha_\mathrm{eff} = \frac{\alpha s_\mathrm{T}}{\delta_\mathrm{s}}$ is the effective Gilbert damping parameter of the system, \(\delta_\mathrm{s} = s_\mathrm{A} - s_\mathrm{B}\) is the net angular momentum, $\gamma_\mathrm{eff}$ is the effective gyromagnetic ratio. As discussed in \cite{diona2025observation}, domain wall dynamics exhibits two distinct saturation mechanisms: at low fields, the competition between SOT and the in-plane magnetic field limits the velocity (Eq.~\ref{vsteadycomplete}), whereas at high current and field the spin wave group velocity $v_{\mathrm{g,max}}$ sets the ultimate bound (Eq.~\ref{rel2}).

\begin{figure*}[t]
    \centering
    \includegraphics[width=0.9\linewidth]{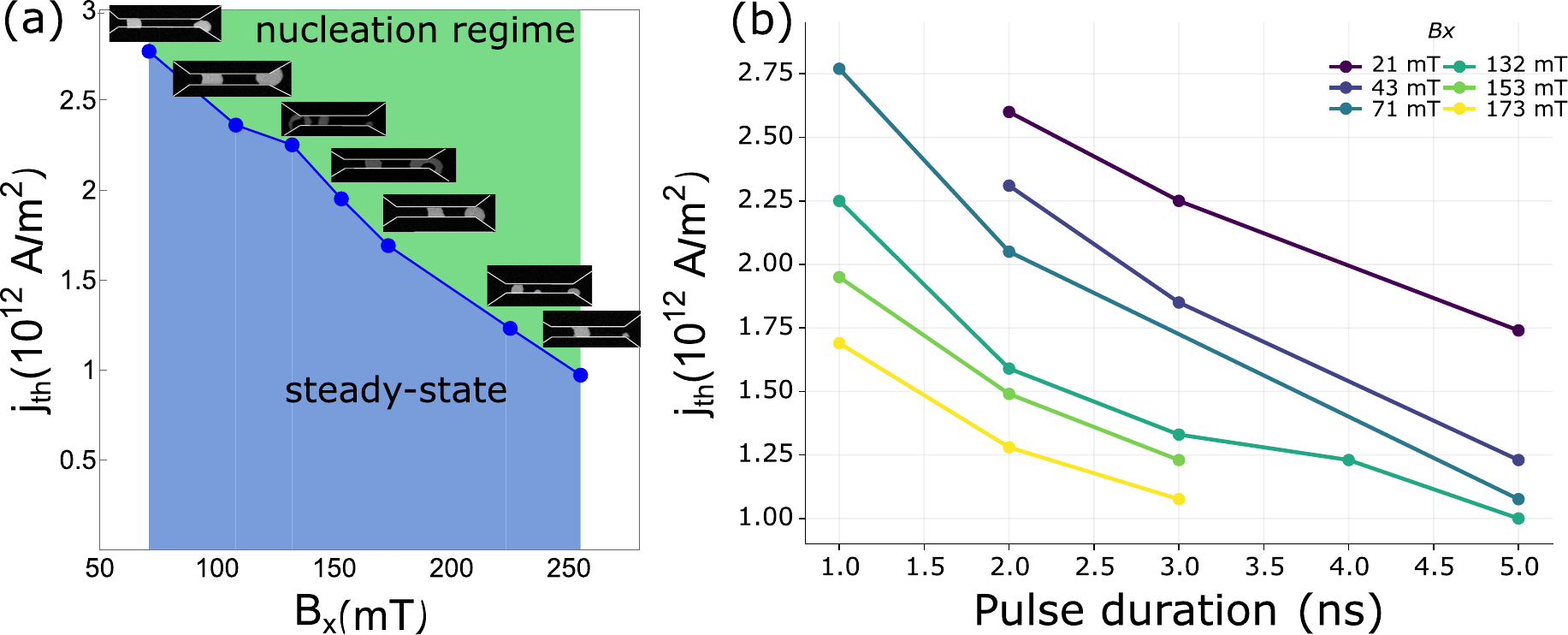}    
    \caption{(a) Nucleation threshold as a function of in-plane magnetic field $B_\mathrm{x}$ and current density $j$ in $\mathrm{Gd}_{27.5}\mathrm{Co}_{72.5}$. Blue markers identify the onset of spontaneous domain nucleation induced by single 1 ns current pulse. Below this boundary, domain wall motion occurs without nucleation, defining a regime of steady-state propagation (light-blue region). Above it, the combined action of field and SOT leads to the formation of additional domains, preventing pure domain wall motion (green region). 
    (b) Threshold current density required for spontaneous domain nucleation as a function of pulse duration at different in-plane magnetic fields in $\mathrm{Gd}_{27.5}\mathrm{Co}_{72.5}$. The threshold current decreases for longer pulses, consistent with thermally assisted nucleation and a time-dependent activation barrier.
    }
    \label{fig2}
\end{figure*}

\begin{figure*}[t]
     \centering
         \includegraphics[width=1\linewidth]{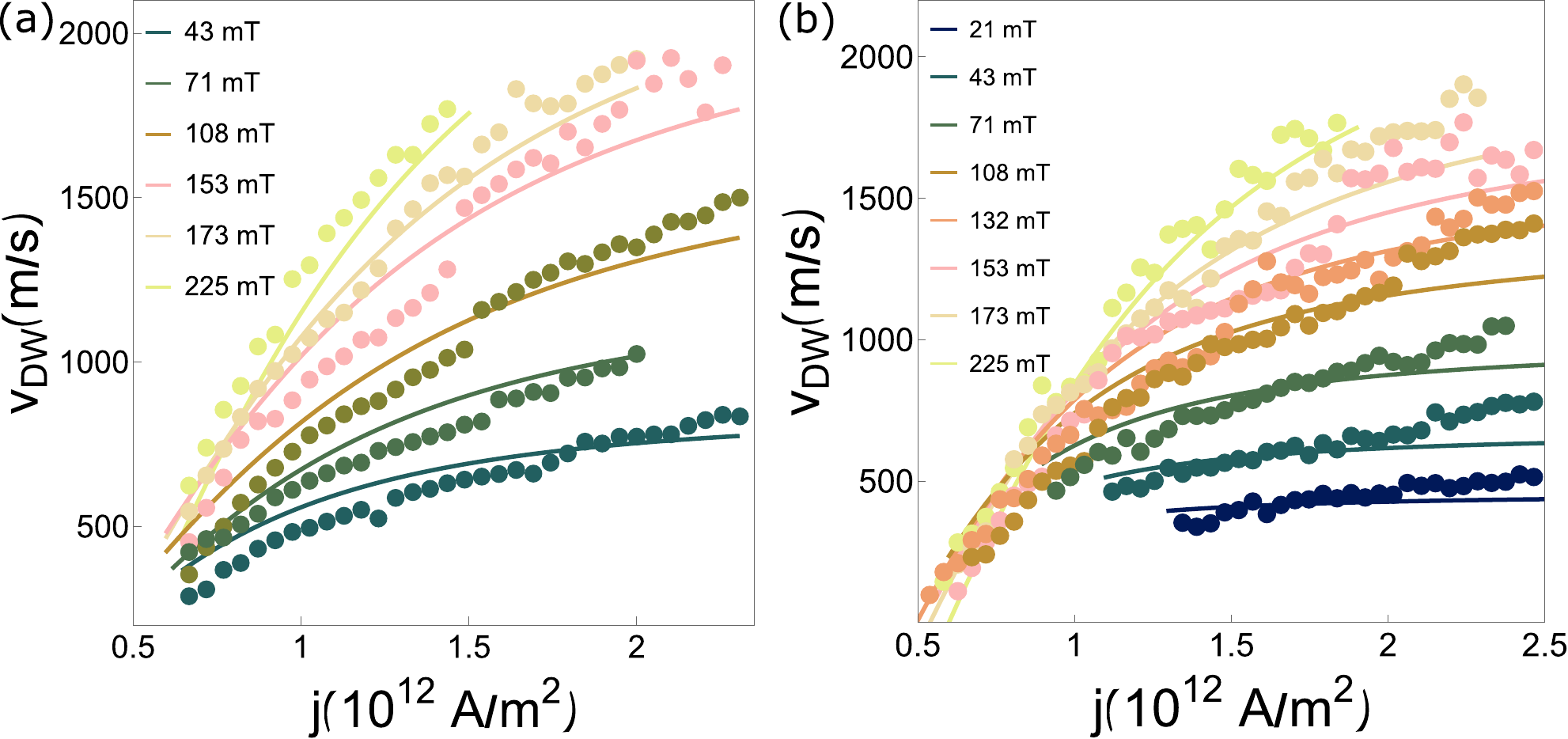}
         
    \caption{Domain wall speed as a function of the applied current density through the heavy metal layer for different in-plane magnetic fields in (a) \(\mathrm{Gd_{27.5}Co_{72.5}}\) and (b) \(\mathrm{Gd_{31.3}Co_{68.7}}\) nanowires. The velocity increases with both in-plane magnetic field and current density, with no saturation or field-independent high-speed collapse. This behavior establishes that GdCo remains outside the relativistic regime before nucleation intervenes; the estimated magnon velocity exceeds $7~\mathrm{km/s}$.}
       \label{fig3}
\end{figure*}

In RE–TM ferrimagnets, this relativistic limit was long considered experimentally inaccessible \cite{caretta2020relativistic}, due to the large antiferromagnetic exchange interaction $A$ compared to ferrimagnetic garnets. However, $A$ depends strongly on growth and processing conditions--including annealing, oxygen content, pressure, and alloy stoichiometry \cite{brunsch1977perpendicular, katayama1978annealing, nishihara1979effects}. Consequently, reported values range from $\approx 1\text{–}2~\mathrm{pJ/m}$ \cite{raasch1994exchange, diona2025observation, sala2023deterministic} to above $10~\mathrm{pJ/m}$ \cite{brunsch1977perpendicular, katayama1978annealing, nishihara1979effects}, resulting in variations of $v_{\mathrm{g,max}}$ of up to an order of magnitude, from $\sim 1.5\text{–}2~\mathrm{km/s}$ to beyond $10~\mathrm{km/s}$.
As a result, access to the relativistic regime is governed by the exchange interaction $A$ and by the stability of the magnetic state under large currents required to approach $v_{\mathrm{g,max}}$. For low $A$, as in GdFeCo, domain walls can approach the magnon velocity before competing instabilities dominate \cite{diona2025observation}. For large $A$, as in the GdCo nanowires studied here, the increased magnon velocity requires current densities and magnetic fields that instead trigger the domain nucleation. The relevant experimental limit is therefore not the intrinsic spin-wave velocity itself, but a nucleation boundary in the space of current density, in-plane field, and pulse duration. We map this boundary by measuring the maximum current density for deterministic propagation as a function of in-plane field using 1 ns pulses (Fig.~\ref{fig2}a). Longer pulses reduce the threshold current (Fig.~\ref{fig2}b), consistent with thermally assisted nucleation in which the applied drive lowers an effective energy barrier and Joule heating increases the probability of barrier crossing \cite{sala2023deterministic}. Consequently, the measured domain wall velocity remains well below $v_{\mathrm{g,max}}$, and the dynamics is classical. Consistently, the velocity increases monotonically with magnetic field (Fig.~\ref{fig3}), in contrast to the relativistic saturation observed in GdFeCo \cite{diona2025observation}. The nucleation boundary therefore defines a thresholded nonlinear activation regime and provides a potential physical basis for future nonlinear spintronic computing architectures.

\paragraph{Experimental discussion --- }\label{sec12}

We measured the domain wall motion in RE-TM ferrimagnetic nanowires with the magneto-optical Kerr effect (MOKE) microscope. Two samples are characterized and more exactly $\mathrm{Si/SiN(8)/Pt(6)/Gd_{x}Co_{1-x}(10)/SiN(8)}$, with $\mathrm{x = 27.5, 31.3}$ (the numbers in parentheses indicate the nanometric thickness of each layer). The two concentrations are found respectively before and after the magnetization compensation point. Consequently, at room temperature, the samples work close to the spin angular momentum compensation point, allowing for maximization of domain wall speed in the range of km/s \cite{caretta2018fast, cai2020ultrafast, kim2017fast, sala2022asynchronous, diona2025observation}. 
The GdCo films were patterned into 40 \textmu m-long nanowires onto Pt layer by a combination of lithography and physical etching techniques. The Pt Hall layer serves as a current pulse injection line to generate the SOT that displaces the domain wall. 

The experiment starts with the nucleation of a domain wall in the nanowire by means of an out-of-plane magnetic field. Then, current pulses are sent through the nanowire to displace the domain wall by SOT. The domain wall displacement is directly probed using MOKE microscopy. In this technique, the polarization of reflected light rotates depending on the local magnetization of the sample, providing magnetic contrast \cite{lo2016magnetic}. Fig.~\ref{fig1} shows representative MOKE images of current-driven domain wall displacement under different in-plane magnetic fields. By quantifying the net displacement of the domain wall $\Delta  x_{\mathrm{DW}}$ and knowing the number and length of current pulses sent through the nanowire, it is possible to extract the velocity of the domain wall. 
Figure~\ref{fig3} shows the domain wall velocity as a function of current density for different in-plane magnetic fields in GdCo nanowires. In contrast to GdFeCo \cite{diona2025observation}, the velocity does not saturate and velocity-current curves for different fields do not collapse onto a single high-field curve. Instead, domain wall velocity increases monotonically with both current density and field over the entire investigated range. This behavior is the central experimental indication that GdCo remains outside the relativistic regime: no field-independent high-speed transport is observed, and the velocity retains a clear dependence on $B_x$. The absence of saturation indicates that substantially larger driving forces would be required to approach the magnon velocity. This points to a higher spin-wave group velocity in GdCo, consistent with a stronger antiferromagnetic exchange interaction $A$ compared to GdFeCo.

To quantify this behavior, we fit the measured domain wall velocity $v_\mathrm{DW}$ using Eq.~\ref{vsteadycomplete}, as shown in Fig.~\ref{fig3}. The analysis confirms that domain walls remain in the classical regime, well below the relativistic limit. This is consistent with the larger antiferromagnetic exchange interaction extracted in our system compared to GdFeCo \cite{diona2025observation}, which results in a higher spin wave velocity, as summarized in Table~\ref{paramextracted}. To limit the number of free parameters, the magnetic anisotropy, saturation magnetization, and SOT efficiency were independently measured via magnetometry and magnetotransport. The remaining parameters are extracted from the fits under the same assumptions adopted in \cite{diona2025observation} (Table~\ref{paramextracted}). These include $A$, $M_\mathrm{A}$, $\alpha$, $H_\mathrm{D}$, and $H_\mathrm{pin}$. The extracted values are consistent with typical RE–TM ferrimagnets \cite{hansen1991magnetic}, and we fix the lattice spacing to $d = 4$~\AA \cite{diona2025observation, caretta2020relativistic}.
The effective DMI field $H_\mathrm{D}$ is extracted by fitting the domain wall velocity curves in the low-current, low-field regime. In this limit, the domain wall speed (Eq.~\ref{vsteadycomplete}) reduces to $v_{\mathrm{cl}} = \gamma_{\mathrm{eff}}\Delta\left(H_D + \frac{\pi}{2}H_\mathrm{x}\right)$, allowing $H_\mathrm{D}$ to be directly determined as the in-plane field approaches zero. The extracted values are reported in Table~\ref{paramextracted} and are in good agreement with previous reports \cite{morshed2021tuning, kim2019bulk}. In contrast to the samples studied in \cite{diona2025observation}, where a negligible DMI field was observed, our GdCo nanowires exhibit a finite $H_\mathrm{D}$. We attribute this difference to the reduced thickness of the ferrimagnetic layer. Here, the GdCo thickness is 10 nm, compared to 15 nm in GdFeCo \cite{diona2025observation}. The smaller thickness enhances interfacial contributions relative to bulk effects, leading to a measurable DMI field.
Fitting the data with Eq.~\ref{vsteadycomplete} yields a $\chi^2$ comparable to that obtained using Eq.~\ref{rel2}. This reflects the fact that, far from the spin wave group velocity, relativistic effects are negligible and no Lorentz-like contraction occurs. In this limit, Eq.~\ref{rel2} reduces to Eq.~\ref{vsteadycomplete}, so both models provide equivalent fits and return similar parameters. As a consequence, the exchange interaction $A$ cannot be accurately determined from these fits, although it is consistently found to be approximately $10~\mathrm{pJ/m}$, in agreement with reported values for GdCo \cite{brunsch1977perpendicular, katayama1978annealing, nishihara1979effects}.
While the large exchange interaction shifts the relativistic regime beyond experimentally accessible driving conditions, the observed nucleation instability introduces a competing dynamical regime. The transition between steady-state propagation and spontaneous domain formation is characterized by a threshold in current density, magnetic field, and pulse duration. Such behavior produces a nonlinear response, where small variations of the excitation parameters can trigger abrupt changes in the magnetic configuration. The decrease of the threshold current with pulse duration indicates an activation process consistent with thermally assisted barrier crossing (Fig.~\ref{fig2}). In an Arrhenius picture, the nucleation probability during a pulse of duration $\tau$ scales as $P = 1 -\tau f_0 \exp\left(-\frac{\Delta E(B_x) - \gamma j}{k_\mathrm{B}T-\eta j^2}\right)$ where $ f_0$ is the characteristic frequency, $\Delta E(B_x)$ is the surface tension of the domain wall, lowered by the SOT current density and the in-plane field; the temperature $T$ increases due to Joule heating \cite{sala2023deterministic}. 
Although the present work does not implement a computing device, these thresholded nucleation dynamics provide a potential physical basis for future nonlinear spintronic computing architectures.

\begin{table}[h!]
 \caption{Parameters of Equation~\ref{vsteadycomplete} for two alloy compositions. A, $M_{\mathrm{A}}$, $\alpha$, $H_{\mathrm{D}}$, $H_{\mathrm{pin}}$ are fitted to the experimental data. $v_{\mathrm{g,max}}$ is derived from the fitted parameters. $K_{\mathrm{eff}}$, $M_{\mathrm{S}}$ and $\theta_{\mathrm{DL}}$ are experimentally measured.}
    \begin{tabular}{@{}lll@{}} 
    \hline
    Parameters & $\mathrm{Gd_{27.5}Co_{72.5}}$ & $\mathrm{Gd_{31.3}Co_{68.7}}$\\
    \hline\hline
    $A$  & $\mathrm{\approx 11\,pJ/m}$ & $\mathrm{\approx 12 \, pJ/m}$\\
    \hline
    $M_\mathrm{A}$ & $\mathrm{1.4\cdot 10^{6} \,A/m}$ & $\mathrm{1.3\cdot 10^{6} \, A/m}$\\
    \hline
    $\alpha$ & 0.0075 & 0.004\\
    \hline
    $H_\mathrm{D}$ & 38  mT & 34 mT\\
    \hline
    $H_\mathrm{pin}$ & 10  mT & 22 mT\\
    \hline
    $v_\mathrm{g,max}$ & $\mathrm{\approx 7.5 \, km/s}$ & $\mathrm{\approx 9 \, km/s}$\\
    \hline\hline
    $K_\mathrm{eff}$ & $\mathrm{3.8\cdot 10^4 \frac{J}{m^3}}$ & $\mathrm{6.3\cdot 10^4 \frac{J}{m^3}}$\\
    \hline
    $M_\mathrm{S}$ & $\mathrm{1.8\cdot 10^5 \frac{A}{m}}$ & $\mathrm{1.7\cdot 10^5 \frac{A}{m}}$\\
    \hline
    $\theta_\mathrm{DL}$ & $0.13$ & $0.13$\\
    \hline
    \end{tabular}
    \label{paramextracted}
\end{table}

\paragraph{Conclusions --- }
We have investigated current-driven domain wall dynamics in GdCo nanowires using MOKE measurements. In contrast to ferrimagnetic systems that access the relativistic regime \cite{caretta2020relativistic, diona2025observation}, in GdCo we find no evidence of velocity saturation. The domain wall velocity increases monotonically with current density and in-plane magnetic field, indicating that the dynamics remain in the classical regime throughout the accessible experimental window. This behavior is attributed to the larger antiferromagnetic exchange interaction in GdCo, which raises the maximum spin-wave group velocity to $\approx 7\text{--}9~\mathrm{km/s}$. Before this intrinsic limit can be approached, increasing current and field promote spontaneous domain nucleation at preferential sites, suppressing steady-state domain wall propagation. This defines a boundary between deterministic transport and a nucleation-dominated regime, setting the limit to the achievable domain wall velocity. Additionally, we observe a finite Dzyaloshinskii--Moriya interaction, attributed to the reduced thickness of the GdCo layer, which enhances interfacial effects with the adjacent Pt layer.
Overall, our results show that relativistic transport and nucleation are competing regimes of ferrimagnetic domain wall dynamics. In low-exchange ferrimagnets such as GdFeCo, the relativistic regime is experimentally accessible; in high-exchange GdCo, thermally assisted nucleation preempts relativistic transport and defines a dynamical phase boundary between steady propagation and nonlinear activation. These findings provide guidelines for designing high-speed ferrimagnetic spintronic devices and identify nucleation as a material-level nonlinear response relevant to future spintronic computing architectures.

\bibliography{bibliography}



\section*{Competing interests}
The author declares no conflict of interest.

\section*{Acknowledgment}
The author acknowledges Dr. Sergey Artyukhin for insightful discussions, valuable feedback on the manuscript, and support in the development and derivation of the analytical model, and Dr. Giacomo Sala for valuable discussions and insightful comments regarding the experimental measurements and data interpretation.

\end{document}